\documentclass[10pt,twoside,twocolumn,english,aps,manuscript,aps,preprint,showpacs,superscriptaddress,showkeys]{revtex4}
\usepackage{lmodern}
\usepackage[T1]{fontenc}
\usepackage[latin9]{inputenc}
\usepackage[active]{srcltx}
\usepackage{graphicx}
\usepackage{setspace}

\makeatletter
\@ifundefined{textcolor}{}
{%
 \definecolor{BLACK}{gray}{0}
 \definecolor{WHITE}{gray}{1}
 \definecolor{RED}{rgb}{1,0,0}
 \definecolor{GREEN}{rgb}{0,1,0}
 \definecolor{BLUE}{rgb}{0,0,1}
 \definecolor{CYAN}{cmyk}{1,0,0,0}
 \definecolor{MAGENTA}{cmyk}{0,1,0,0}
 \definecolor{YELLOW}{cmyk}{0,0,1,0}
}

\makeatother

\usepackage{babel}
\begin{document}

\title{Unveiling of Bragg glass to vortex glass transition by an ac driving
force in a single crystal of Yb$_{3}$Rh$_{4}$Sn$_{13}$}

\author{Santosh Kumar}

\email{santoshkumar@phy.iitb.ac.in}

\address{Department of Physics, Indian Institute of Technology Bombay, Mumbai
400076, India.}

\author{Ravi P. Singh}

\altaffiliation{Present address: Department of Physics, Indian Institute of Science Education and Research Bhopal, Bhopal 462066, India.}

\address{Department of Condensed Matter Physics and Materials Science, Tata
Institute of Fundamental Research, Mumbai 400005, India.}

\author{A. Thamizhavel}

\address{Department of Condensed Matter Physics and Materials Science, Tata
Institute of Fundamental Research, Mumbai 400005, India.}

\author{C. V. Tomy}

\address{Department of Physics, Indian Institute of Technology Bombay, Mumbai
400076, India.}

\author{A. K. Grover}

\email{arunkgrover@gmail.com}

\address{Department of Condensed Matter Physics and Materials Science, Tata
Institute of Fundamental Research, Mumbai 400005, India.}

\address{Department of Physics, Panjab University, Chandigarh 160014, India.}
\begin{abstract}
We present here some striking discrepancies in the results of ac and
dc magnetization measurements performed in a single crystal of low
$T_{c}$ superconductor, Yb$_{3}$Rh$_{4}$Sn$_{13}$. Fingerprint
of a transition from an ordered vortex lattice \textit{a la} Bragg
glass (BG) phase to a partially-disordered vortex glass (VG) like
phase gets unearthed under the influence of an ac driving force present
inevitably in the isothermal ac susceptibility ($\chi\thinspace^{\prime}(H)$)
measurements. In contrast to its well\,-\,known effect of improving
the state of spatial order in the vortex matter, the ac drive is surprisingly
found to promote disorder by assisting the BG to VG transition to
occur at a lower field value in this compound. On the other hand,
the isothermal dc magnetization ($M$--$H$) scans, devoid of such
a driving force, do not reveal this transition; they instead yield
signature of another order-disorder transition at elevated fields,
viz., peak effect (PE), located substantially above the BG to VG transition
observed in $\chi\thinspace^{\prime}(H)$ runs. Further, the evolution
of PE feature with increasing field as observed in isofield ac susceptibility
($\chi\thinspace^{\prime}(T)$) plots indicates emergence of an ordered
vortex configuration (BG) from a disordered phase above a certain
field, $H^{*}$\,($\sim4$\,kOe). Below $H^{*}$, the vortex matter
created via field\,-\,cooling (FC) is found to be better spatially
ordered than that prepared in zero field\,-\,cooled (ZFC) mode.
This is contrary to the usual behavior anticipated near the high-field
order-disorder transition (PE) wherein a FC state is supposed to be
a supercooled disordered phase and the ZFC state is comparatively
better ordered.
\end{abstract}



\maketitle

\section{Introduction}

In the context of the mixed state of a type\,-\,II superconductor,
the seminal discovery \cite{key-1,key-2} of a well-ordered thermodynamic
phase, viz., Bragg glass (BG) exhibiting Bragg\textquoteright s reflections,
and its possible transition(s) \cite{key-3} to a disordered phase
devoid of Bragg's reflections had lead to a generic vortex phase diagram
\cite{key-4} applicable to almost all pinned superconductors. Owing
to the possibility of a sudden proliferation of dislocations on progressively
increasing the magnetic field at a constant temperature, the quasi\,-\,long
range ordered BG phase is anticipated \cite{key-1,key-2} to transform
first into a multi\,-\,domain (partially disordered) vortex glass
(VG) phase. Such a transition usually reflects as a second peak in
(isothermal) magnetization ($M$--$H$) loops, termed as the second
magnetization peak (SMP) anomaly \cite{key-5,key-6,key-7,key-8,key-9,key-10,key-11}.
Thereafter, at elevated fields closer to the upper critical field
($H_{c2}(T)$), there occurs another anomaly, known as the quintessential
peak effect (PE) phenomenon \cite{key-3,key-9,key-10,key-12,key-13,key-14,key-15,key-16,key-17,key-18}
in field/temperature variation in critical current density, $j_{c}(H,T)$.
The PE is argued \cite{key-12} to signal the collapse of the elasticity
of an ordered vortex lattice at a rate faster than the pinning force
density near $H_{c2}(T)$. 

Although theoretical treatment related to the BG to VG transition
exists in the literature (for example, as in Ref.~\cite{key-2}),
however, the experimental tools, particularly those employed to explore
the bulk pinning properties, such as, the dc magnetization ($M$--$H$)
measurements do not always capture this transition. Therefore, in
the context of vortex phase diagram studies, it is tempting to ask
a question- \textit{Is the BG to VG phase transition generic?} If
it is so, then this anomaly must get exposed in the $H$--$T$ space
of all pinned superconductors, possessing a certain amount of quenched
disorder. To address this issue, we have investigated via detailed
magnetization measurements, a low $T_{c}$ superconductor Yb$_{3}$Rh$_{4}$Sn$_{13}$,
which has so far been reported \cite{key-14,key-15,key-16} to display
only the PE phenomenon. A claim of the presence of SMP anomaly (or
its counterpart) in this compound has so far not been made by anyone.
We have now found that the BG to VG (akin to SMP anomaly) transition
in this compound gets exposed prior to the onset of PE under the influence
of an ac driving force present in the (isothermal) ac susceptibility
scans ($\chi\thinspace^{\prime}(H)$). Counter-intuitively, shaking
of the vortex array by an ac drive in the present study has been seen
to promote the spatial disordering in the vortex matter by assisting
the BG to VG transition process. This observation is in complete contrast
to the usual role of an ac driving force, which is to improve the
state of spatial order in multi\,-\,domain vortex matter as reflected
\cite{key-17} by the enhanced brightness of the Bragg spots in the
field-temperature phase space prior to crossover to the PE region
in the vortex phase diagram. Another interesting aspect of the present
study is a revelation of the inequality $j{}_{c}^{FC}(H)<j{}_{c}^{ZFC}(H)$
at lower fields (below a characteristic field, $H^{*}\approx4$\,kOe)
. This amounts to stating that the vortex matter created in the field\,-\,cooled
mode (FC) below $H^{*}$ exhibits better spatial ordering than that
obtained in the zero field\,-\,cooled manner. This feature also
is in sharp contrast to the behavior \cite{key-19,key-20,key-21,key-22}
seen at the higher fields, where one encounters order\,-\,disorder
transition (\textit{a la} PE phenomenon). The vortex matter created
via field\,-\,cooling is expected to be spatially more disordered
(than the ZFC state) due to supercooling of an amorphous vortex matter
below the PE region. We present evidences that $H^{*}$ signifies
a crossover regime from an ordered BG phase into a disordered amorphous
phase while reducing the field.

\section{Experimental details}

Single crystals of Yb$_{3}$Rh$_{4}$Sn$_{13}$ have been grown by
tin flux method \cite{key-23}. The specimen chosen for the present
study is platelet shaped, with a planar area of $5.76$\,mm$^{2}$
and thickness of $0.62$\,mm. The superconducting transition temperature
($T_{c}$) of this crystal is found to be nearly $7.5$\,K. Magnetization
data, both ac as well as dc, were recorded using the same instrument,
viz., Superconducting Quantum Interference Device\,-\,Vibrating
Sample Magnetometer (SQUID\,-\,VSM, Quantum Design Inc., USA). The
magnetic field was directed along the crystalline {[}110{]} axis,
with a possible error in the alignment to be within $5^{0}$. The
demagnetization factor in this orientation is expected to be small
as the field is applied in the plane of the thin platelet, i.e., normal
to the smallest dimension (thickness) of the sample. In the dc measurements,
we kept the amplitude of vibration of the sample to be small ($\approx0.5$\,mm)
so as to minimize the field-inhomogeneity along the scan length. During
the ac susceptibility measurements ($\chi\thinspace^{\prime}(H,T)$),
an ac field of amplitude $1$\,Oe and frequency $211$\,Hz was superimposed
on the applied dc field. 
\begin{figure}[t]
\begin{singlespace}
\begin{centering}
\includegraphics[scale=0.33]{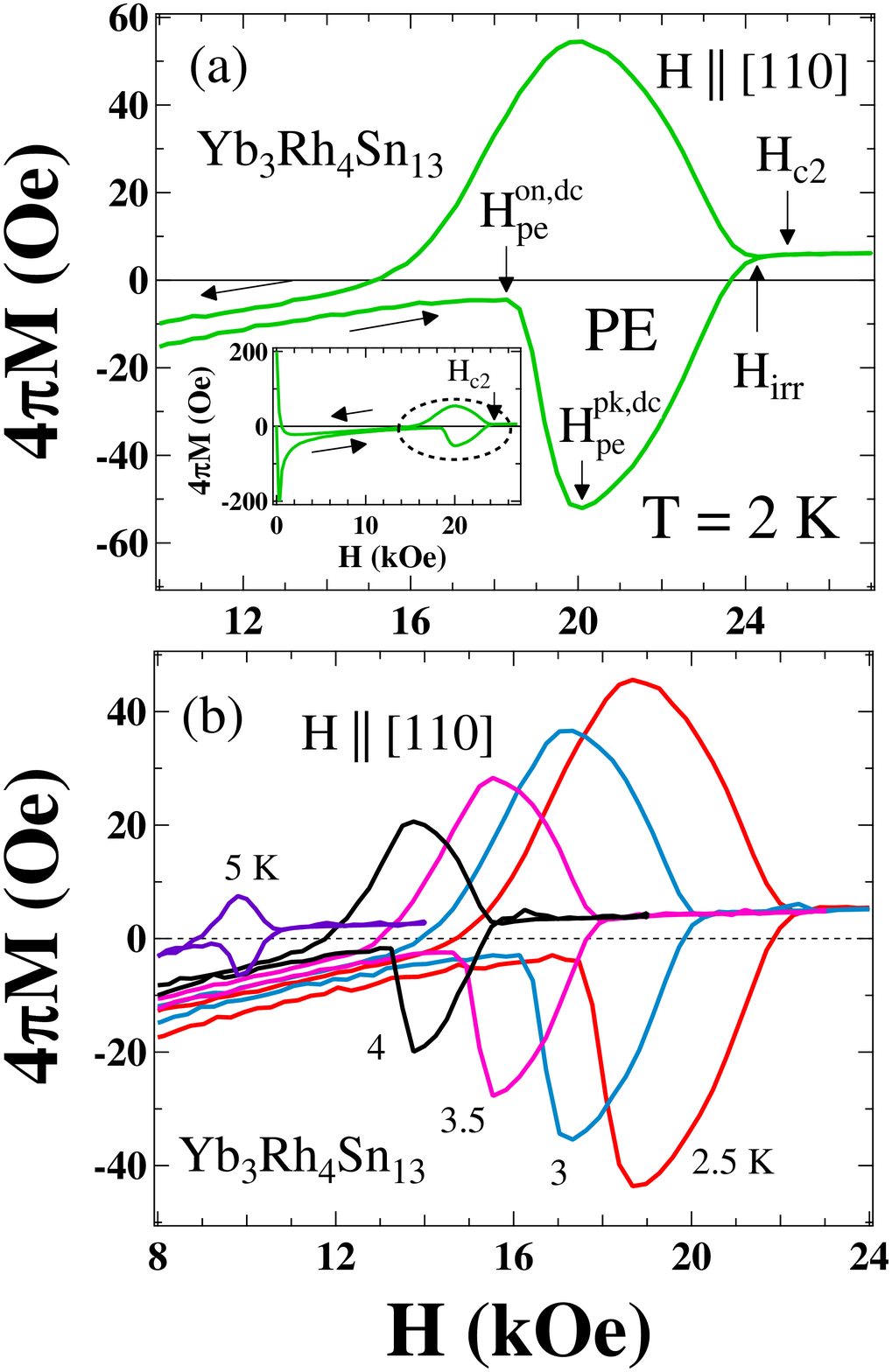}
\par\end{centering}
\end{singlespace}

\begin{singlespace}
\protect\caption{(Color online) (a) An expanded portion of (isothermal) dc $M$--$H$
loop at $T=2$\,K demonstrating a typical PE phenomenon in a single
crystal of Yb$_{3}$Rh$_{4}$Sn$_{13}$. The inset displays the first
two quadrants of this loop in the entire field range of investigation,
i.e., $0<H<28$\,kOe. (b) The $M$--$H$ curves at different temperatures
show the evolution of the PE feature as it occurs at different fields. }
\end{singlespace}
\end{figure}

\section{Results}

\subsection{Isothermal dc $M$--$H$ loops: Manifestation of the peak effect
phenomenon}

The inset panel of Fig.~1(a) displays the first two quadrants of
an isothermal dc magnetization hysteresis loop ($M$--$H$) obtained
at $T=2$\,K for field applied parallel to the {[}110{]} plane of
Yb$_{3}$Rh$_{4}$Sn$_{13}$ crystal. The $M(H)$ curve can be seen
to be hysteretic between the forward ($M(H^{+})$) and the reverse
($M(H^{-})$) sweeps of the magnetic field, as expected for a pinned
type\,-\,II superconductor. However, there exists an unusual enhancement
in the hysteresis width ($\Delta M(H)=M(H^{+})-M(H^{-})$) prior to
$H_{c2}$, as apparent from the encircled portion of the $M(H)$ curve.
A magnified view of this portion is displayed on an expanded scale
in the main panel of Fig.~1(a). Using a prescription of the Bean\textquoteright s
Critical State Model \cite{key-24}, Fietz and Webb have shown \cite{key-25}
that the hysteresis width ($\Delta M$) can be taken as a measure
of the critical current density, $j_{c}(H,T)$. Therefore, the enhancement
in $\Delta M$ reflects an unusual increase in $j_{c}(H)$ a little
below $H_{c2}$, which can be identified as the PE phenomenon. The
onset field ($H_{pe}^{on,dc}$) and the peak field ($H_{pe}^{pk,dc}$)
of the PE stand located in the main panel of Fig.~1(a). The merger
of $M(H^{+})$ and $M(H^{-})$ beyond the bubble feature identifies
the irreversibility field ($H_{irr}$).

We show in Fig.~1(b), the fingerprint of PE feature at different
temperatures, as indicated. On increasing the temperature, a systematic
decrease in the onset and the peak field values of the PE is well
apparent. The decrease in $H_{pe}^{on,dc}$ and $H_{pe}^{pk,dc}$
with $T$, nearly following the variation of $H_{c2}$ with $T$,
is a characteristic feature of the PE phenomenon. The order\,-\,disorder
transition pertaining to PE has been argued to have first\,-\,order
character, as was well evident in the results of small angle neutron
scattering study in a low $T_{c}$ superconductor Nb \cite{key-17}
and those of scanning Hall probe microscopy in 2H-NbSe$_{2}$ \cite{key-26}.
\begin{figure}[b]
\begin{centering}
\includegraphics[scale=0.33]{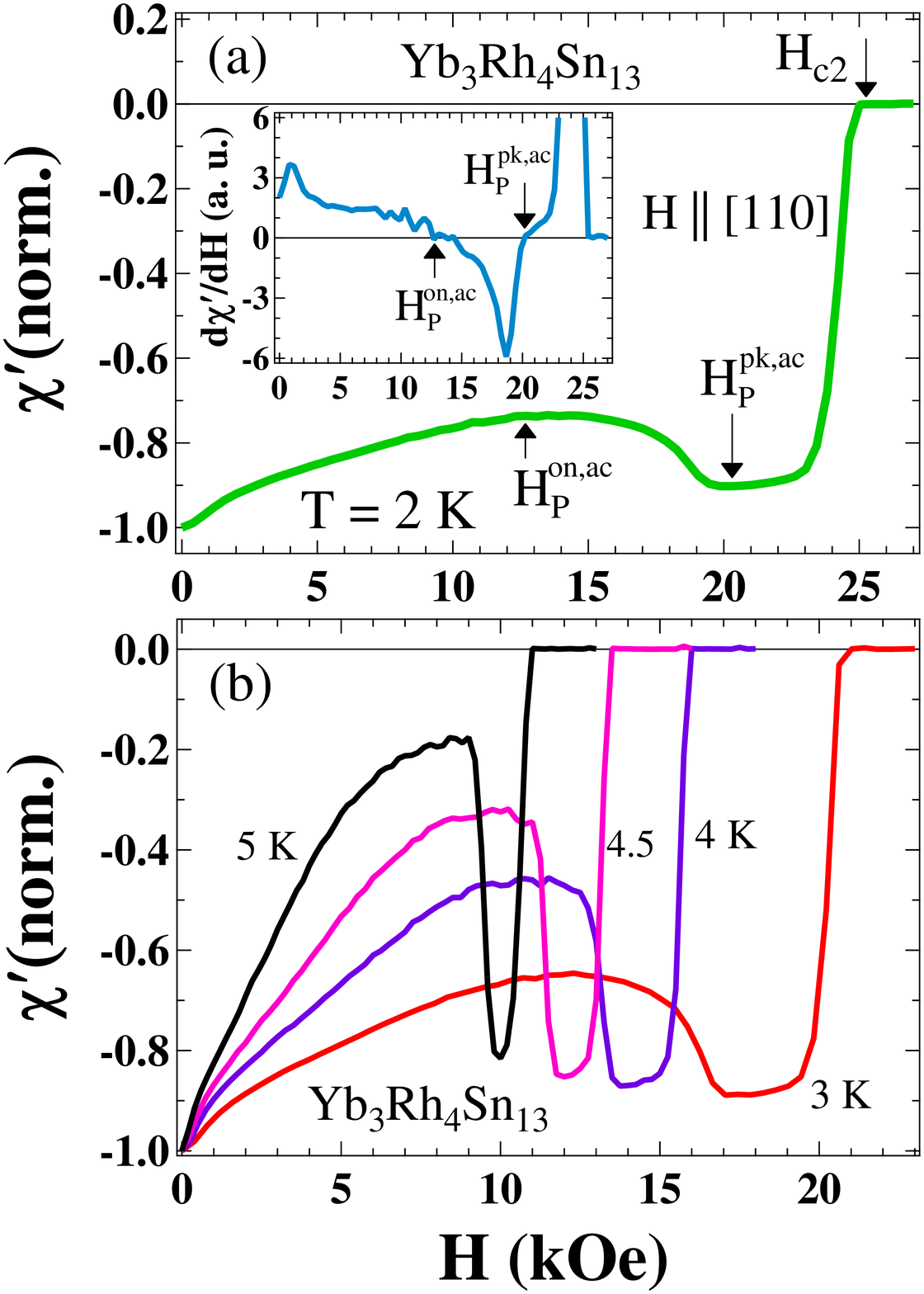}
\par\end{centering}

\protect\caption{(Color online) (a) The real part of (isothermal) ac susceptibility
($\chi\thinspace^{\prime}(H)$) data plotted against the dc magnetic
field at $T=2$\,K. A broad dip\,-\,like anomaly triggering at
a field marked as $H_{p}^{on,ac}$ has been observed. The inset panel
illustrates a portion of $d\chi\thinspace^{\prime}/dH$ curve at $T=2$\,K
which helps to identify the onset ($H_{p}^{on,ac}$) and peak ($H_{p}^{pk,ac}$)
positions of the anomaly identified respectively via the first and
second zero\,-\,crossings of the derivative plot. (b) $\chi\thinspace^{\prime}(H)$
curves at different temperatures show how the anomaly changes from
a broad to a narrow one with increase in temperature. }
\end{figure}

\subsection{Isothermal ac susceptibility $\chi\thinspace^{\prime}(H)$ responses:
Identification of an additional anomaly in conjunction with the PE
phenomenon}

To investigate further the order\,-\,disorder transition(s) in the
vortex matter, we recorded the isothermal ac susceptibility ($\chi\thinspace^{\prime}(H)$)
responses as shown in Fig.~2. The sample was initially cooled\,-\,down
to a chosen temperature in (near) zero field and thereafter the $\chi\thinspace^{\prime}$
data were recorded while ramping the field to higher values. The |$\chi\thinspace^{\prime}$|
values at $T=2$\,K can be seen to fall monotonically with increasing
field until about a certain field, marked as $H_{p}^{on,ac}$\,($\approx12.5$\,kOe,
cf.~main panel of Fig.~2(a)). Above $H_{p}^{on,ac}$, a broad (anomalous)
dip\,-\,like feature encompassing a large field interval (i.e.,
$H_{p}^{on,ac}<H<H_{c2}$) can be noticed. Note that the identification
of onset field ($H_{p}^{on,ac}$) of this dip feature was made possible
by locating the first zero\,-\,crossing in the derivative plot of
$\chi\thinspace^{\prime}(H)$ as displayed in an inset panel of Fig.~2(a).
The second zero\,-\,crossing in $d\chi\thinspace^{\prime}/dH$ identifies
the peak position (marked as $H_{p}^{pk,ac}$) of the anomalous variation
in $\chi\thinspace^{\prime}(H)$ across the dip region. We recall
here that the $\chi\thinspace^{\prime}$ value reflects $j_{c}$ through
the two relations \cite{key-27}, (i) $\chi\thinspace^{\prime}\sim-1+\alpha h_{ac}/j_{c}$
for $h_{ac}<h^{*}$ and (ii) $\chi\thinspace^{\prime}\sim-\beta j_{c}/h_{ac}$
if $h_{ac}>h^{*}$, where $\alpha$ and $\beta$ are size and geometry
dependent factor, $h_{ac}$ is the amplitude of the ac field and $h^{*}$
is the ac field for full penetration. Following equation (i), the
anomalous variation in $\chi\thinspace^{\prime}(H)$ above $H_{p}^{on,ac}$
in Fig.~2(a) amounts to an unusual increase in otherwise monotonically
decreasing $j_{c}(H)$. Similar anomalous behaviour in $\chi\thinspace^{\prime}(H)$
can also be observed at different temperatures as illustrated in Fig.~2(b).
While the dip feature remains quite broad at lower temperatures, one
can note that the width of this anomalous region gets substantially
reduced on higher temperature side. For example, the dip in $\chi\thinspace^{\prime}(H)$
at $5$\,K is observed to be much sharper with a narrow transition
width (of nearly $2$\,kOe) prior to approaching $H_{c2}$ which,
following equation (ii), amounts to a rapid increase in $j_{c}(H)$
echoing the characteristics of the quintessential PE phenomenon.

A comparison of $\chi\thinspace^{\prime}(H)$ data presented in Fig.~2(a)
with that of the $M$--$H$ loop shown in Fig.~1(a), tells us that
the peak field value ($H_{p}^{pk,ac}\approx20$\,kOe) of the anomaly
observed in the former is nearly the same as the corresponding peak
field value ($H_{pe}^{pk,dc}$) of the PE observed at the same temperature
in the latter. However, if we seek the analog of onset field of the
anomaly seen in ac $\chi\thinspace^{\prime}(H)$, viz., $H_{p}^{on,ac}$\,($\approx12.5$\,kOe,
Fig.~2(a)), we find that no anomalous feature stands depicted at
this field value in the dc $M$--$H$ loop (cf.~Fig.~1(a)). This
implies an additional anomaly has got unearthed (at $H_{p}^{on,ac}$)
in the $\chi\thinspace^{\prime}(H)$ scans in the specimen of Yb$_{3}$Rh$_{4}$Sn$_{13}$
which is not observed in its $M$--$H$ runs. To seek the rationalization
of this discrepancy in the ac and dc magnetization data, one can consider
the presence of an ac driving force inevitably superimposed on the
dc field during the $\chi\thinspace^{\prime}(H)$ runs as an important
difference between the two sets of measurements. Two important factors
in conjunction, viz., (i) the shaking effect of an ac driving force
on the vortex matter and, (ii) driving effect due to a continuous
ramping of the dc magnetic field, may be acting as a combined driving
force leading to the triggering of additional anomaly in $\chi\thinspace^{\prime}(H)$
runs. To shed some more light onto this assertion, it became tempting
to explore the isofield ac susceptibility responses, where an effect
due to the ramping of magnetic field would not be present. These are
described ahead. 
\begin{figure}[t]
\begin{centering}
\includegraphics[scale=0.37]{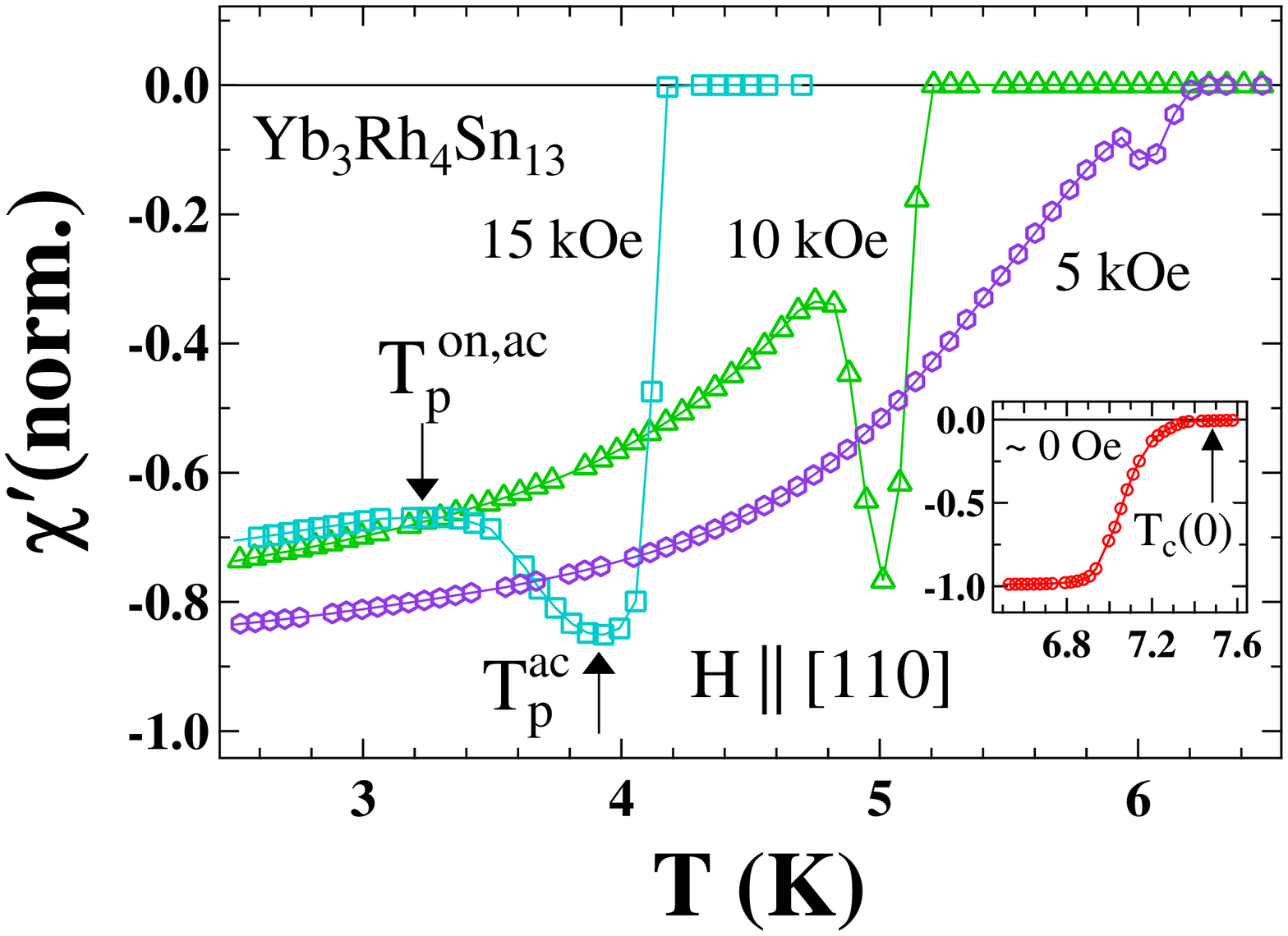}
\par\end{centering}

\protect\caption{(Color online) Observation of PE anomaly in temperature\,-\,dependent
ac susceptibility ($\chi\thinspace^{\prime}(T)$) runs at various
fixed dc fields. Onset ($T_{p}^{on,ac}$) and peak ($T_{p}^{ac}$)
temperature of PE have been marked in the $\chi\thinspace^{\prime}(T)$
curve for $H_{dc}=15$\,kOe. The inset shows the occurrence of zero\,-\,field
superconducting\,-\,normal transition at $T\approx7.5$\,K. }
\end{figure}

\subsection{Isofield ac susceptibility $\chi\thinspace^{\prime}(T)$ responses:
identification of the PE phenomenon}

Figure~3 displays the temperature dependences of the in\,-\,phase
ac susceptibility ($\chi\thinspace^{\prime}(T)$) obtained at various
fixed dc fields. The sample was initially cooled down to $1.8$\,K
in (nominal) zero field, a desired field was then applied and the
$\chi\thinspace^{\prime}(T)$ data were recorded while warming up
to higher temperatures ($T>T_{c}$). The zero\,-\,field superconducting
transition temperature, $T_{c}(0)$ identified via the onset of diamagnetic
response is found to be about $7.5$\,K (cf. inset panel of Fig.~3).
Fingerprint of the PE feature identified by a dip\,-\,like characteristic
can be observed in all the curves shown in the main panel of Fig.~3.
At $H=5$\,kOe, the dip in $\chi\thinspace^{\prime}(T)$ is found
to be less prominent, however, it evolves into a sharp (negative)
peak at higher field values (see, e.g., $\chi\thinspace^{\prime}(T)$
response in $H=10$\,kOe). On further increasing the field to $15$\,kOe,
the PE region becomes broad, which is apparent from a wider gap between
the onset ($T_{p}^{on,ac}$) and the peak ($T_{p}^{ac}$) temperatures
of the PE marked in the main panel of Fig.~3. Such a broadening seen
in the PE feature at higher fields may be ascribed to effects of an
enhancement in \textit{'effective pinning'} with field, as articulated
by Giamarchi and Le Doussal \cite{key-1,key-2} and elucidated in
a crystal of low $T_{c}$ superconductor 2H\,-\,NbSe$_{2}$ by Banerjee
\textit{et al.}, \cite{key-21}. 

A lesser developed (PE) anomaly at lower field value ($\sim5$\,kOe)
suggests the possibility of a lesser ordered vortex matter emerging
at that end as well. Such a trend is consistent with the notion that
an ordered vortex lattice (Bragg glass) undergoes a transition into
a disordered phase on lowering the field \cite{key-28,key-29}. 

Across the field interval from about $5$\,kOe to $14$\,kOe, the
vortex matter seems to evolve into a better spatially ordered (Bragg
glass) phase, as is evidenced by a sharp PE feature at $10$\,kOe
in Fig.~3.
\begin{figure}[tp]
\begin{centering}
\includegraphics[scale=0.55]{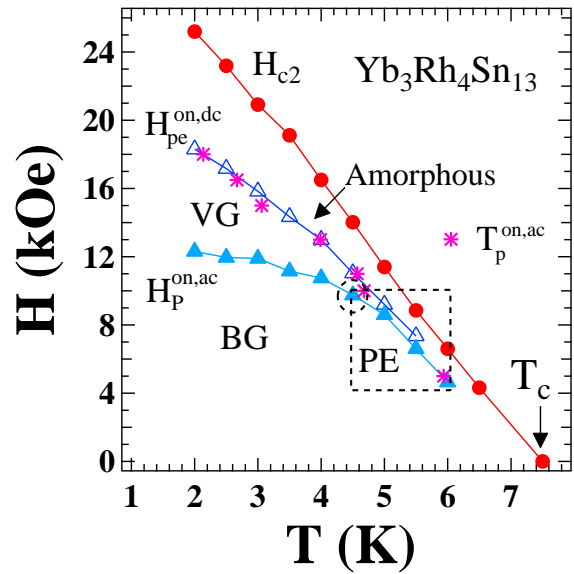}
\par\end{centering}

\protect\caption{(Color online) A sketch of vortex phase diagram in our crystal of
Yb$_{3}$Rh$_{4}$Sn$_{13}$. $H_{p}^{on,ac}(T)$ and $H_{pe}^{on,dc}(T)$
have been extracted from $\chi\thinspace^{\prime}(H)$ and $M$--$H$
plots, respectively. Below $T=4.5$\,K, the former resembles the
BG to VG transition line (refer text in section\,-\,4, \textquotedblleft Discussion\textquotedblright )
while the latter portrays the characteristics of onset of the PE anomaly.
In the boxed\,-\,region, both $H_{p}^{on,ac}(T)$ and $H_{pe}^{on,dc}(T)$
lines behave as the onset of the PE transition. The onset temperatures
($T_{p}^{on,ac}(H)$) of the PE obtained from isofield $\chi\thinspace^{\prime}(T)$
scans almost fall on the onset field values $H_{pe}^{on,dc}(T)$ of
the PE extracted from $M$--$H$ loops. $H_{c2}(T)$ taken from $M$--$H$
loops depicts usual linear fall with increase in temperature. }
\end{figure}

\subsection{$H$--$T$ phase diagram}

It is instructive to compare the results of ac and dc magnetization
data of Figs.~1 to 3. For this purpose, we present in Fig.~4, a
$H$--$T$ phase diagram of Yb$_{3}$Rh$_{4}$Sn$_{13}$, which comprises
the field/temperature values corresponding to the onset positions
of the anomalies seen in Figs.~1 to 3. The following features in
this phase diagram are noteworthy:

1. The onset field values ($H_{pe}^{on,dc}(T)$) of the PE (open triangles)
acquired from the $M$--$H$ data (Fig.~1) fall smoothly with the
increase in temperature, a trend similar to the $H_{c2}(T)$ line. 

2. The corresponding onset field values ($H_{p}^{on,ac}(T)$ shown
by closed triangles) of the anomaly observed in $\chi\thinspace^{\prime}(H)$
plots (Fig.~2), however, do not coincide with the onset position
($H_{pe}^{on,dc}(T)$) of the PE over a significantly large portion
of the $H$--$T$ space (i.e., $T<4.5$\,K). Here, the $H_{p}^{on,ac}(T)$
line stays substantially below the $H_{pe}^{on,dc}(T)$ line. Moreover,
$H_{p}^{on,ac}(T)$ values exhibit a weaker temperature dependence
at temperatures below $4.5$\,K unlike the faster temperature dependence
seen in case of $H_{pe}^{on,dc}(T)$. The two onset field lines, $H_{pe}^{on,dc}(T)$
and $H_{p}^{on,ac}(T)$, come closer to each other only in the boxed
region (i.e., at $T>4.5$\,K and $5$\,kOe\,$<H<10$\,kOe) and
show identical temperature dependence. 

3. The discrepancy in the phase diagram (Fig.~4) pertaining to the
location of onset fields of the anomalies seen in the (isothermal)
$M$--$H$ and $\chi\thinspace^{\prime}(H)$ data prompts the need
to take into consideration the results of isofield $\chi\thinspace^{\prime}(T)$
scans (Fig.~3) as well. For this purpose, we have plotted in the
phase diagram the onset temperatures ($T_{p}^{on,ac}(H)$) of the
PE (shown by stars) obtained from $\chi\thinspace^{\prime}(T)$ scans.
It is curious to note that the $T_{p}^{on,ac}(H)$ values fall almost
on the onset field ($H_{pe}^{on,dc}(T)$) line of the PE transition.
\textit{We draw an important inference here that the results of isothermal
dc $M$--$H$ loops and the isofield ac }$\chi\thinspace^{\prime}(T)$\textit{
scans predict the onset of the PE anomaly at nearly the same phase
boundary, i.e., ($H_{pe}^{on,dc}$, $T_{p}^{on,ac}$) line.} 

Clearly, there exists an additional anomaly (apart from the PE) located
deeper (at $H_{p}^{on,ac}(T)$) in the mixed state of Yb$_{3}$Rh$_{4}$Sn$_{13}$,
which has got unveiled only from the outcomes of the $\chi\thinspace^{\prime}(H)$
data. The other two measurement techniques (viz., the $M$--$H$ and
$\chi\thinspace^{\prime}(T)$ scans), on the other hand, reveal the
fingerprints of only one kind of anomaly, i.e., the PE. This corroborates
our previous assertion that both an ac driving force as well as the
dc magnetic field ramping together trigger the said additional anomaly
well below the onset of PE (these two factors not being present together
during the individual dc $M$--$H$ and ac $\chi\thinspace^{\prime}(T)$
runs). 

The lowest field down to which the PE could be discernible in Fig.~4
is nearly $4.5$\,kOe. Since there is hardly any fingerprint of PE
feature below this value, we could surmise that the vortex matter
there may be disordered. This proposition is explored further via
the thermomagnetic history\,-\,dependent magnetization measurements.
\begin{figure}[tp]
\begin{centering}
\includegraphics[scale=0.38]{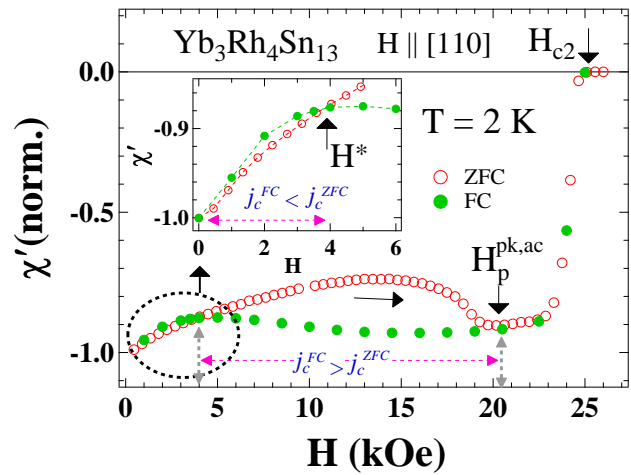}
\par\end{centering}

\protect\caption{(Color online) A comparison of $\chi\thinspace^{\prime}(H)$ plots
recorded in ZFC and FC modes at $T=2$\,K. Across the field range
$4$\,kOe\,$<H<H_{p}^{pk,ac}$, the FC state is more diamagnetic
than the ZFC, implying the inequality, $j_{c}^{FC}>j_{c}^{ZFC}$.
A reversal of this situation, i.e., $j_{c}^{FC}<j_{c}^{ZFC}$ is observed
below $4$\,kOe ($\sim H^{*}$) in an expanded portion of $\chi\thinspace^{\prime}(H)$
as shown in the inset panel. }
\end{figure}

\subsection{Thermomagnetic history dependence in $j_{c}(H,T)$: FC state is more
ordered than the ZFC at low fields}

Figure~5 displays the shielding responses ($\chi\thinspace^{\prime}(H)$)
obtained at $2$\,K for the system prepared in two different histories,
viz., zero field\,-\,cooled (ZFC) and field\,-\,cooled (FC) modes.
The isothermal $\chi\thinspace^{\prime}(H)$ data in the ZFC mode
(open circles) are the same as that presented in Fig.~2(a). In the
FC case, the sample was first cooled from normal state ($T>T_{c}(0)$),
in the presence of a certain applied field ($H<H_{c2}$) down to $T=2$\,K
and a given $\chi\thinspace^{\prime}(H)$ value was recorded. Thereafter,
the sample was warmed up again to a higher $T$\,($>T_{c}(0)$) and
the same procedure was followed for recording another $\chi\thinspace{}_{FC}^{\prime}$
value in a different cooling field. A collation of such $\chi\thinspace{}_{FC}^{\prime}(H)$
data at $2$\,K is illustrated (closed circles) in Fig.~5. Three
different field intervals can be identified; (I) Across the range,
$4$\,kOe\,$<H<H_{p}^{pk,ac}$, the ZFC and FC curves can be seen
to be well separated with the latter possessing more diamagnetic values
than the former. Following equations (Ref.~\cite{key-27}) (i) $\chi\thinspace^{\prime}\sim-1+\alpha h_{ac}/j_{c}$
and (ii) $\chi\thinspace^{\prime}\sim-\beta j_{c}/h_{ac}$, a more
negative $\chi\thinspace^{\prime}$ implies larger $j_{c}$. Therefore,
in the field interval $4$\,kOe\,$<H<H_{p}^{pk,ac}$, the FC state
exhibits a higher $j_{c}$ than that in the ZFC state (i.e., $j_{c}^{FC}>j_{c}^{ZFC}$).
As per a description of Larkin\,-\,Ovchinnikov collective pinning
theory \cite{key-30,key-31} for weakly-pinned superconductors, $j_{c}$
relates inversely to the volume ($V_{c}$) of a domain within which
the vortices are collectively pinned and remain well correlated. Therefore,
a higher $j_{c}$ amounts to a smaller $V_{c}$, and hence signifies
a more (strongly\,-\,pinned) disordered vortex matter. (II) The
history\,-\,dependence in $j_{c}(H)$ tends to cease above $H_{p}^{pk,ac}$
as we observe the two curves to overlap there. This is in line with
the understanding that the vortex matter above the peak field of the
PE is generally believed to be \textit{'disordered in equilibrium'}
\cite{key-17}. (III) The ZFC and FC curves appear to overlap at low
fields ($H<4$\,kOe) as well, as apparent from the two sets of data
points in the encircled portion of $\chi\thinspace^{\prime}(H)$ (cf.
main panel of Fig.~5). However, a closer examination of the two $\chi\thinspace^{\prime}(H)$
curves in this region on an expanded scale (see inset panel of Fig.~5)
reveals slightly less diamagnetic values (at a given $H$) for the
FC state than that for the ZFC mode, which implies the reversal of
the above inequality; i.e., $j_{c}^{FC}<j_{c}^{ZFC}$ for $H<4$\,kOe.
We have marked this crossover field value ($\sim4$\,kOe) as $H^{*}$
in the inset panel of Fig.~5. Such an inequality ($j_{c}^{FC}<j_{c}^{ZFC}$)
below $H^{*}$ is found to be very robust as it can be observed at
other temperatures as well (all data not shown here but the $H^{*}$
values at different temperatures have been displayed in the $H$--$T$
phase diagram as shown ahead in Fig.~8). 
\begin{figure}[tp]
\begin{centering}
\includegraphics[scale=0.35]{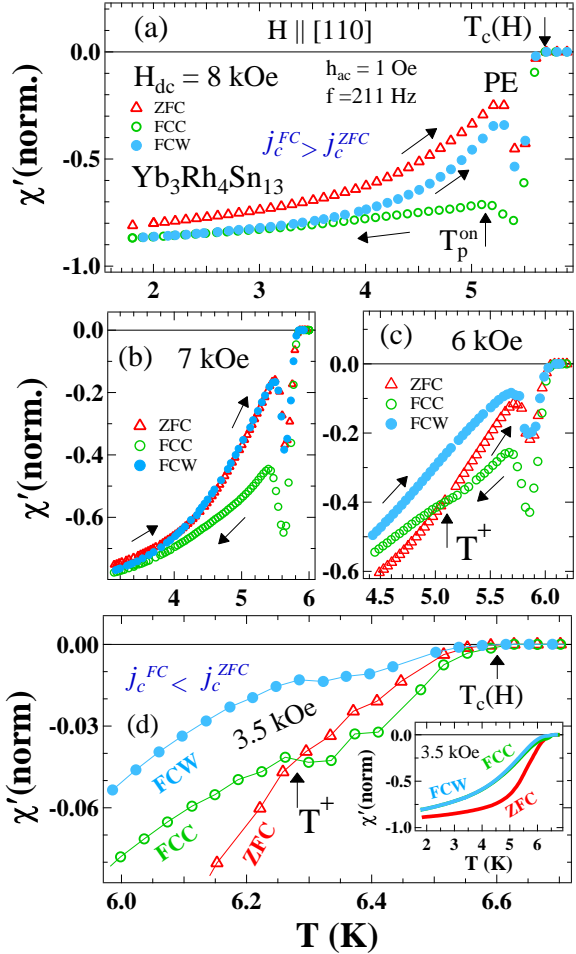}
\par\end{centering}

\protect\caption{(Color online) Temperature\,-\,dependent $\chi\thinspace^{\prime}(T)$
curves obtained in ZFC, FCC and FCW modes at $H_{dc}=$ (a) $8$\,kOe,
(b) $7$\,kOe, (c) $6$\,kOe and (d) $3.5$\,kOe. At higher fields
($8$\,kOe, panel (a)), the FC curves (FCC and FCW) are more diamagnetic
than the ZFC and hence, they imply the inequality $j_{c}^{FC}>j_{c}^{ZFC}$.
On reducing the field to $6$\,kOe (panel (c)) and then to $3.5$\,kOe
(inset panel of (d)), the ZFC curve becomes the most diamagnetic below
a characteristic temperature $T^{+}$ (i.e., $j_{c}^{FC}<j_{c}^{ZFC}$).
At $H=3.5$\,kOe (see main panel of (d)) there is no signature of
PE in the ZFC mode while a residual fingerprint of it can be seen
in the FC curves. }
\end{figure}

The $H$--$T$ phase space region below $H^{*}$ was further explored
via isofield $\chi\thinspace^{\prime}(T)$ scans recorded in different
modes, viz., zero field\,-\,cooled (ZFC), field\,-\,cooled cool\,-\,down
(FCC) and field\,-\,cooled warm\,-\,up (FCW) runs as depicted
in Fig.~6. The sample was cooled in (near) zero field from the normal
state down to $1.8$\,K, and then a desired magnetic field was applied.
Thereafter, the $\chi\thinspace^{\prime}(T)$ data were obtained (marked
as ZFC in Fig.~6), while warming the sample to higher $T$. Following
this, $\chi\thinspace^{\prime}(T)$ data were again recorded while
cooling the sample down to $1.8$\,K (marked as FCC) in the presence
of the same applied field, and thereafter while warming it towards
the normal state (marked as FCW). At $H=8$\,kOe (Fig.~6(a)), the
shielding response is found to be less diamagnetic for the ZFC case
than that observed for the FCC and FCW modes (i.e., $j_{c}^{FC}(8\thinspace kOe)>j_{c}^{ZFC}(8\thinspace kOe)$
at all temperatures). This implies that the vortex matter created
in the ZFC mode is better spatially ordered among the three modes
which is consistent with the observations made at this field value
in Fig.~5. At a lower field ($H=7$\,kOe), the ZFC curve nearly
overlaps with the FCW; these two curves stay less diamagnetic than
the FCC for $T>4$\,K (cf.~Fig.~6(b)). On reducing the field further
to $6$\,kOe (Fig.~6(c)), the FCW curve can now be seen to be the
least diamagnetic amongst the three curves at all temperatures. Further,
there is an unusual intersection of ZFC and FCC curves at a certain
temperature marked as $T\,^{+}$, below which the ZFC is now more
diamagnetic than the FCC and FCW curves. The ZFC curve at a lower
field of $3.5$\,kOe (see inset panel of Fig.~6(d)) can be seen
to be more diamagnetic and well separated from the other two (FCC
and FCW) curves. This would imply the inequality $j_{c}^{FC}<j_{c}^{ZFC}$,
which, in turn, suggests that the vortex matter created in the ZFC
mode is more disordered than that created in the FC mode. This situation
is the reversal of that depicted in Fig.~6(a) and is in agreement
with the conclusion drawn from the history\,-\,dependent $\chi\thinspace^{\prime}(H)$
responses below $H^{*}\approx4$\,kOe (cf.~Fig.~5). An expanded
portion of $\chi\thinspace^{\prime}(T)$ plots at $H=3.5$\,kOe in
the main panel of Fig.~6(d) shows that the fingerprint of PE ceases
to exists in the ZFC mode, whereas a tiny modulation can still be
noticed across $T\,^{+}<T<T_{c}(H)$ in the case of FCC and FCW runs.
This suggests that the vortex state created in ZFC manner at $H=3.5$\,kOe
is disordered to such an extent that any order\,-\,disorder vortex
phase transition may not be identifiable in the temperature-dependent
warm-up measurements. On the other hand, at the same field value ($3.5$\,kOe),
a nascent signature \textit{a la} PE observed in FCC and FCW modes
suggests that vortex matter is comparatively better ordered when created
in the FC mode at $3.5$\,kOe. 
\begin{figure}[t]
\begin{centering}
\includegraphics[scale=0.5]{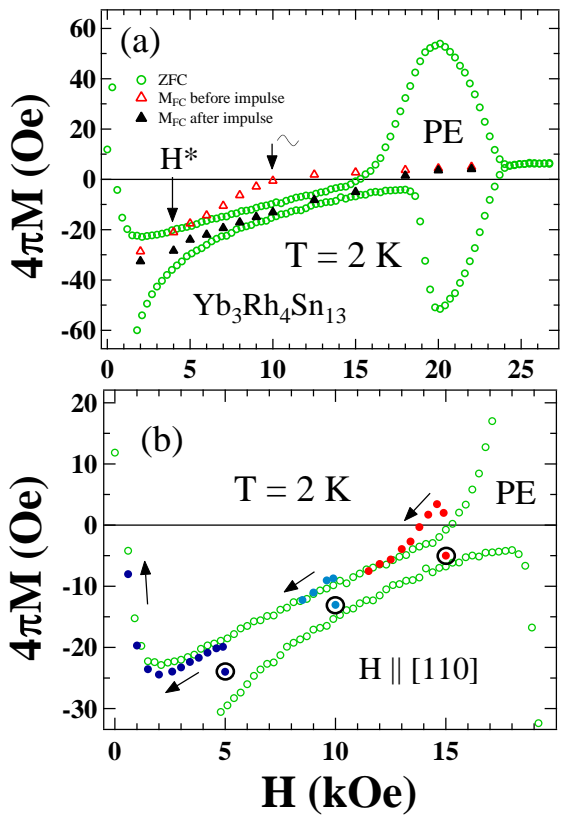}
\par\end{centering}

\protect\caption{(Color online) (a) $M$--$H$ loop at $T=2$\,K in the ZFC mode (open
circles). The field-cooled magnetization values ($M_{FC}$, open triangles)
for $5$\,kOe\,$<H<15$\,kOe unexpectedly lay outside the ZFC envelope
loop. An impulse of an ac field (amplitude\,$=10$\,Oe, $f=211$\,Hz)
imposed on the FC states created at different $H$ (open triangles)
leads the $M_{FC}$ values to fall well within the loop (closed triangles).
(b) Minor hysteresis loops at $T=2$\,K traced while reducing the
field to zero after creating the FC states at $H=5$, $10$, $15$\,kOe
and imposing an ac field impulse on them. }
\end{figure}

Another demonstration of a more ordered FC state (than the ZFC state)
at lower field values is apparent from the results of thermomagnetic
history\,-\,dependent dc $M$--$H$ measurements, as illustrated
in Fig.~7. The $M$--$H$ curve shown by open circles pertains to
the usual hysteresis loop obtained in the ZFC mode at $T=2$\,K.
Magnetization data recorded in the FC mode ($M_{FC}$), i.e., after
cooling the sample from a higher $T$ ($>T_{c}(0)$) down to $2$\,K
in the presence of different chosen field values have also been displayed
by open triangle data points in Fig.~7(a). If we associate the $M_{FC}$
values to the equilibrium (reversible) magnetization ($M_{eq}$) and
ignore the contribution due to bulk currents that can get set up due
to a gradient in macroscopic field ($H$) (as in the case of ZFC),
then, as per equation $M_{eq}=(M(H^{+})+M(H^{-}))/2$ \cite{key-25},
the $M_{FC}$ data are supposed to fall in the middle of the ZFC envelope
loop (as, for example, evident in Ref.~\cite{key-10}). Figure~7(a)
depicts an unexpected scenario, wherein the $M_{FC}$ data points
stay outside the envelope hysteresis loop for field ($H$) values
in the interval, $5$\,kOe\,$<H<15$\,kOe. We can try to rationalize
this using a model due to Clem and Hao \cite{key-32} which accounts
for the FC magnetization of a type\,-\,II superconductor when a
gradient in macroscopic field gets established as a consequence of
flux expulsion during the FC mode. In such a situation, the $M_{FC}$
values deviate from $M_{eq}$ (ideal case of no flux\,-\,pinning).
This deviation of FC magnetization from $M_{eq}$ values is indeed
anticipated to be governed by the strength of bulk pinning at a given
field; the extent of deviation is more (less) for stronger (weaker)
pinning strength ($j_{c}$) (see Fig.~5 in Ref.~\cite{key-32}).
The $M_{FC}$ values (see open triangles, Fig.~7(a)) falling outside
the envelope loop in the present case, for $H>5$\,kOe are due to
a larger $j_{c}$ value associated with the vortex matter created
in the FC mode and the flux density gradient \cite{key-32} that gets
set-up while field\,-\,cooling. Note that below $H=5$\,kOe (close
to $H^{*}\sim4.5$\,kOe), the FC state seems to have a lower $j_{c}$
value (better ordered) as the $M_{FC}$ value here remains well within
the envelope (i.e., nearer to $M_{eq}$). 

As a further experimentation, on each occasion, after a FC state was
created at a chosen field value, we momentarily perturbed it with
an ac field impulse of amplitude $10$\,Oe at a frequency of $211$\,Hz
(applied for about $6$\,seconds). It is curious to note that the
magnetization recorded after the impulse treatment (closed triangles
data points in Fig. 7(a)) fall well within the envelope loop. It appears
as if the magnetization response of a FC state after perturbation
by an impulse conforms to the anticipated respective equilibrium value
($M_{eq}$) (closed triangles data points) located in the middle of
the ZFC envelope loop. The imposition of an ac field impulse results
in the reconfiguration of an unperturbed FC state into state, comparison
of whose $j_{c}$ value with that of the corresponding ZFC state can
be very instructive. We thus recorded the 'FC minor hysteresis loops'.
Figure~7(b) shows the magnetization curve obtained while reducing
the dc field to zero after the (perturbed) FC states were created
(encircled) at $H=5$\,kOe, $10$\,kOe and $15$\,kOe. The initial
magnetization value recorded while decreasing the field from $H=5$\,kOe
undershoots the ZFC envelope, and the magnetization curve thereafter
traverses a path that remains within the envelope loop as the field
is ramped down to the zero value. Taking cue from the linear relation
between hysteresis width and $j_{c}$ \cite{key-25}, the (perturbed)
FC state is reckoned to have a lower $j_{c}$ value than that in the
ZFC for fields below $H^{*}$. This observation fortifies the inferences
drawn from data in Figs.~5 and 6. At $H=10$\,kOe, the magnetization
data, while reducing the field, first overshoots the ZFC envelope
though marginally, and thereafter it retraces the reverse leg ($M(H^{-})$)
of the ZFC magnetization curve. It can be argued that the FC state
here exhibits $j_{c}$ value slightly higher than that in the corresponding
ZFC states. A significant overshooting of the ZFC envelope loop is
witnessed at a higher field of $15$\,kOe, which indicates a larger
$j_{c}$ value for the FC state than that created in the ZFC mode.
Note that the field value of $15$\,kOe lies in the anomalous region,
$H_{p}^{on,ac}(T)<H<H_{pe}^{on,dc}(T)$ of the $H$--$T$ space at
$2$\,K (Fig.~4) which, as discussed in section\,-\,IV ahead,
is \textit{'partially disordered in equilibrium'} as the multi-domain
VG state.

\section{Discussion}

The magnetization measurement techniques (both ac and dc) employed
in the present work have lead to new revelations in Yb$_{3}$Rh$_{4}$Sn$_{13}$
which have been summarized in the form of a modified $H$--$T$ phase
diagram in Fig.~8. These include, (i) the occurrence of a broad anomaly
located deeper in the mixed state triggering at $H_{p}^{on,ac}(T)$
line as seen in the $\chi\thinspace^{\prime}(H)$ curves (Fig. 2),
(ii) identification of onset position of the PE at $H_{pe}^{on,dc},\thinspace T{}_{p}^{on,ac}$
line as obtained from both the dc $M$--$H$ (Fig.~1) and the ac
$\chi\thinspace^{\prime}(T)$ (Fig.~3) plots, and (iii) a characteristic
line $H^{*}(T)$ extracted from Fig.~5, below which a vortex matter
created in the FC mode is somewhat more ordered than that obtained
in the ZFC manner. Following a description of L\,-\,O theory \cite{key-30,key-31},
we argue that the anomaly in $\chi\thinspace^{\prime}(H)$ (Fig.~2)
reflects a shrinkage in $V_{c}$ as $j_{c}\propto1/\surd V_{c}$ and
hence, can be termed as an order\,-\,disorder transition in the
vortex matter. Note that the L\,-\,O theory is well applicable for
weakly\,-\,pinned superconductors as in the present case of Yb$_{3}$Rh$_{4}$Sn$_{13}$
(ratio of depinning and depairing current densities is of the order
of $10^{-4}$). The location of $H_{p}^{on,ac}(T)$ line deeper in
the mixed state and its weak temperature\,-\,dependence outside
the boxed region in Fig.~4 can be argued to relate to the disorder\,-\,induced
transition \cite{key-1,key-2,key-6,key-7} of a quasi-ordered (elastic)
vortex lattice \textit{a la} BG phase to the dislocation-mediated
(multi\,-\,domain) VG phase. Note that the peak field of the anomalous
variation in $j_{c}(H)$ as reflected in dc $M$--$H$ and ac $\chi\thinspace^{\prime}(H)$
plots, viz., $H_{pe}^{pk,dc}(T)$ (closed squares) and $H_{p}^{pk,ac}(T)$
(open squares), is consistent, as the two sets of data points extracted
from two different measurements can be seen to be overlapping in Fig.~8.
This would imply that the complete amorphization of the vortex matter
(which is generally believed to occur at peak field of the PE \cite{key-21})
occurs at the same field/temperature value while performing measurements
via ac and dc magnetization techniques.
\begin{figure}[tp]
\begin{centering}
\includegraphics[scale=0.4]{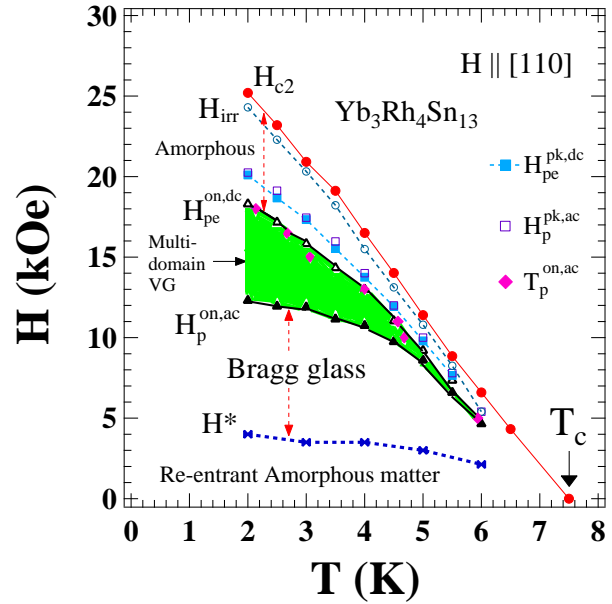}
\par\end{centering}

\protect\caption{(Color online) Complete vortex phase diagram in our crystal of Yb$_{3}$Rh$_{4}$Sn$_{13}$.
Various regions have been identified (refer text for details). A delineation
between the BG to VG transition (occurring at $H_{p}^{on,ac}(T)$)
and the PE anomaly (at $H_{pe}^{on,dc}(T)$), is well apparent. The
peak field values $H_{pe}^{pk,dc}(T)$ and $H_{p}^{pk,ac}(T)$ obtained
respectively from the $M$--$H$ and $\chi\thinspace^{\prime}(H)$
data almost fall on the same phase boundary. The region between ($H_{pe}^{pk,dc}(T)$/$H_{p}^{pk,ac}(T)$)
and the $H_{irr}(T)$ line presumably comprises pinned amorphous matter
while the narrow space between $H_{irr}(T)$ and $H_{c2}(T)$ may
involve an unpinned amorphous vortex matter. Below $H^{*}(T)$ (obtained
from Fig.~5), the $H$--$T$ phase space region comprises (reentrant)
disordered vortex matter, a vortex configuration created here in the
FC manner is found to be slightly more ordered than that in the ZFC
mode.}
\end{figure}

It has been shown earlier that the annealing effects on the vortex
matter, produced either by an ac driving force \cite{key-17,key-33}
or by repeated cycling of the dc magnetic field \cite{key-34,key-35},
eliminate either partly or completely a (metastable) disordered vortex
phase yielding an ordered vortex configuration. Also, the residual
presence of a disordered (metastable) vortex phase usually governs
the onset position of the order-disorder (PE) transition(s) \cite{key-21,key-29,key-36}.
As a usual behavior, the onset position of the PE shifts to higher
field values after the annealing of (disordered) vortex phase, whereas
the same moves to lower fields when the amount of quenched disorder
is larger \cite{key-21,key-29}. Surprisingly, the shaking effect
of an ac driving force appears to be counterintuitive in the present
study, as it promotes the spatial disordering of the vortex matter
rather than its usual role of improving the state of spatial order.
This is well apparent by the (early) occurrence of an order\,-\,disorder
transition at $H_{p}^{on,ac}(T)$ line (cf. Fig. 8) in the $\chi\thinspace^{\prime}(H)$
runs which involve the ac drive. In the absence of an ac driving force
as in dc $M$--$H$ scans, the onset of order-disorder transition
(PE) is seen at higher fields; $H_{pe}^{on,dc}(T)$ values in Fig.~8
are located significantly above the BG to VG transition ($H_{p}^{on,ac}(T)$)
line. If one assumes that the shaking effect on the vortex array by
an ac drive results in the lowest equilibrium-like state of the system
(as in Ref.~\cite{key-17}) under given circumstances, then the region
bounded by the phase lines, $H_{p}^{on,ac}(T)$ and $H_{pe}^{on,dc}(T)$,
in Fig.~8 can be accepted as \textit{``disordered in equilibrium''}
in the form of multi-domain vortex glass phase. The present findings
echo similar assertions made recently by us \cite{key-18} in another
study in a single crystal of a low $T_{c}$ superconductor, Ca$_{3}$Ir$_{4}$Sn$_{13}$.
However, in that study in Ca$_{3}$Ir$_{4}$Sn$_{13}$, a distinct
demarcation between the BG to VG transition line and the locus of
the onset of the PE anomaly in its vortex phase diagram was not apparent,
which has now been clearly sorted out in the phase diagram of Yb$_{3}$Rh$_{4}$Sn$_{13}$.

The nature of BG to VG transition had been argued to be of first\,-\,order
\cite{key-37} in a high $T_{c}$ superconductor. To fortify this
proposition, we may emphasize that the region bounded between $H_{p}^{on,ac}(T)$
and $H_{pe}^{on,dc}(T)$ in Fig.~8 which is shown from the $\chi\thinspace^{\prime}(H)$
data to be disordered in equilibrium is indeed an ordered vortex phase
when viewed from the outcomes of the dc $M$--$H$ measurements. Therefore,
vortex phase in this region can be treated as a superheated ordered
BG phase. This attests the first\,-\,order nature of the BG to VG
transition in a low $T_{c}$ superconductor.
\begin{figure}[tp]
\begin{centering}
\includegraphics[scale=0.45]{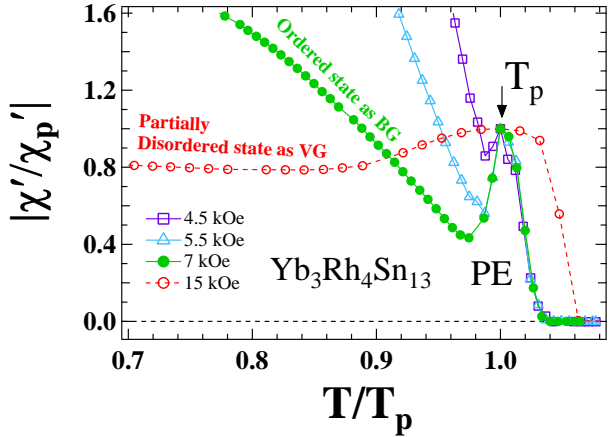}
\par\end{centering}

\protect\caption{(Color online) |$\chi\thinspace^{\prime}/\chi\thinspace{}_{p}^{\prime}$|
data (normalized with respect to $\chi\thinspace^{\prime}$ value
at peak position of the PE) plotted against the reduced temperature,
$T/T_{p}$ at various constant dc fields. The shape of the PE anomaly
changes with field. At $H=4.5$\,kOe, there is a tiny peak feature
while a much prominent peak feature can be noticed at a higher field,
$H=7$\,kOe. The PE feature gets broadened at a further elevated
field of $15$\,kOe.}
\end{figure}

The change in magnetic field tunes the inter-vortex spacing which,
in turn, can influence the balance between the strength of vortex
pinning and the (elastic) interactions between the vortices. As a
result, the extent of spatial ordering/disordering in the vortex matter
may vary in different regions of the $H$--$T$ phase space. Further,
an enhancement in effective pinning at a given field value can give
rise to a qualitative change in the evolution of the size of the Larkin
domain \cite{key-30,key-31}, as had been experimentally demonstrated
\cite{key-38} via the evolution of PE feature in the isofield $\chi\thinspace^{\prime}(T)$
measurements in crystal(s) of 2H-NbSe$_{2}$. Motivated by this, we
shall now examine how the PE feature evolves with magnetic field in
the case of our crystal of Yb$_{3}$Rh$_{4}$Sn$_{13}$. We show in
Fig.~9 the normalized |$\chi\thinspace^{\prime}/\chi\thinspace_{p}^{\prime}$|
plots ($\chi\thinspace{}_{p}^{\prime}$ corresponds to $\chi\thinspace{}^{\prime}$
value at peak position of PE where vortex matter is most disordered
for a given inter-vortex spacing and the underlying quenched random
disorder) against the reduced temperature\,$=T/T_{p}$, ($T_{p}$
is peak temperature of PE obtained in the ZFC mode for various fixed
fields). The PE feature remains absent at fields below $H=4.5$\,kOe
(data not shown here), which may be because of a disordered vortex
matter prevailing there. Note that the larger inter-vortex spacings
at low flux density results in weaker interaction between the vortices
to counter disordering influence of the pinning centres. The vortex
configuration created at low fields ought to be highly disordered
akin to the reentrant amorphous vortex matter anticipated in earlier
studies \cite{key-39,key-40,key-41,key-42}. With the increase in
field, the interaction effects strengthen and consequently, a sparse
disordered vortex configuration can progressively change into an ordered
one, as has been demonstrated earlier in Refs.~\cite{key-43,key-44}.
This is apparent from Fig.~9 by the appearance of a tiny PE feature
firstly at $H=4.5$\,kOe. A more pronounced signature of PE occurring
at $H=5.5$\,kOe suggests further improvement in the spatial ordering
of vortex matter with the increase in field prior to the onset of
PE. At a higher field of $7$\,kOe, there emerges a very well developed
PE feature, which is indicative of a well-ordered vortex (BG) phase
prevailing prior to the onset of the PE transition at this field value.
A different scenario is, however, depicted at a higher field, $H=15$\,kOe,
wherein the order-disorder transition can be seen to be much broader.
We surmise that the vortex phase prior to the order-disorder transition
at this field value is partially disordered (\textit{a la} multi-domain
VG phase) so that any order-disorder transition would reflect a lesser
shrinkage in $V_{c}$ (as compared to that at lower fields) or equivalently,
lesser increment in $j_{c}$ \cite{key-30,key-31} associated with
this state. In all, the |$\chi\thinspace^{\prime}/\chi\thinspace_{p}^{\prime}$|
plots of Fig. 9 have lead us to infer that a disordered vortex phase
at low field end ($H<4.5$\,kOe) transforms into an ordered one (BG
phase) with increase in field. A further enhancement in field may
transform an ordered vortex phase (BG) into a partially disordered
(VG) state prior to the onset of PE.

It is useful to recall now a recent study \cite{key-45} employing
the small angle neutron scattering (SANS) technique in another crystal
of Yb$_{3}$Rh$_{4}$Sn$_{13}$ (which would typically have different
amount of quenched random disorder) providing evidences at microscopic
level in support of our experimental results. In the SANS study, the
evolution of spatial order in vortex matter with vortex lattice spacing
($a_{0}$) has been well-depicted by plots of field-dependences of
the normalized correlation lengths (both longitudinal, $\xi_{L}/a_{0}$
and transversal, $\xi_{T}/a_{0}$) as well as their ratio ($\xi_{L}/\xi_{T}$)
(see Fig.~3 in Ref. \cite{key-45}). These lengths relate to the
longitudinal and radial dimensions of the volume within which the
flux lines remain correlated. The full width at half maximum of the
rocking curves in longitudinal and transverse directions reflecting
the apparent quality of the spatial order and that of the Bragg peaks
remains nearly flat in the field interval $5\thinspace kOe<H<16\thinspace kOe$
(cf. Fig.~2 in Ref. \cite{key-45}). The normalized transversal correlation
length ($\xi_{T}/a_{0}$) starts decreasing more rapidly as the field
reduces below $4$\,kOe \cite{key-45}. Near $4.5$\,kOe, where
we just start to observe the PE feature in |$\chi\thinspace^{\prime}(T)$|
scans in our sample, the said ratio has a value $\sim2$ \cite{key-45}.
This could imply that when the quality of spatial order in the vortex
matter gets more compromised (i.e., $\xi_{T}/a_{0}<2$), the PE feature
would not surface up. The longitudinal correlation length was seen
\cite{key-45} to be nearly fifty times longer than $\xi_{T}$. It
has been argued \cite{key-45} that the ratio $\xi_{L}/\xi_{T}$ reveals
field-induced changes in the parameter $\sqrt{c_{44}/c_{66}}$, where
$c_{44}$ and $c_{66}$ are respectively, the tilt and shear modulii
of hexagonal vortex lattice. This ratio in their sample remains nearly
constant between $5$\,kOe and $12.5$\,kOe and starts to fall above
it and the rate of decrease becomes very steep above $17$\,kOe (see
inset of Fig. 3(a) of Ref. \cite{key-45}). Such a trend could be
interpreted that the quality of spatial order undergoes a change above
about $12.5$\,kOe and the elasticity of the vortex lattice plummets
above $17.5$\,kOe. In the context of our sample, this could rationalize
that BG to VG transition happens above $12.5$\,kOe (see Fig.~8).
The inset panel in Fig.~3(b) of Ref. \cite{key-45} shows that the
simulated (normalized) vortex-vortex interaction in their sample does
not continue to monotonically rise with magnetic field as the inter-vortex
spacing progressively decreases with increase in field. The vortex-vortex
interaction (for the Yb$_{3}$Rh$_{4}$Sn$_{13}$ crystal investigated
in Ref. \cite{key-45}) reaches a maximum value near $12$\,kOe and
starts to decrease above about $14$\,kOe thereby implying that the
onset of this decrease is a consequence of notion of enhancement in
effective pinning at large field values. Near $20$\,kOe, the value
of the normalized vortex-vortex interaction force for this compound
is same as at about $6$\,kOe \cite{key-45}. The rapid decrease
in this parameter below $5$\,kOe reflects the dominance of pinning
in the field domain of reentrant amorphous state (cf. Fig.~8).

In our present study, the |$\chi\thinspace^{\prime}/\chi\thinspace_{p}^{\prime}$|
plots of Fig.~9 summarize the evolution in the spatial order of vortex
matter with magnetic field, which ties up well with the results of
SANS study \cite{key-45}. The extent of spatial ordering/disordering
in the vortex matter at a given field can be quantified by the ratio
of |$\chi\thinspace^{\prime}/\chi\thinspace_{p}^{\prime}$| value
just prior to the PE and that at the peak position (see $T_{p}$ in
Fig.~9) of the PE. Note that the correlation length ought to be minimum
at the peak ($T_{p}$) of the PE and maximum (for a well ordered vortex
lattice) just before the PE. In Fig.~9, the said ratio of |$\chi\thinspace^{\prime}/\chi\thinspace_{p}^{\prime}$|
at onset and peak positions of the PE can be seen to be increasing
smoothly with increasing magnetic field for $4.5\thinspace kOe<H<7\thinspace kOe$,
which is consistent with the variation of correlation lengths with
field for $H<10$\,kOe, as anticipated in Ref. \cite{key-45}. This
would imply that a very well-ordered BG phase emerges nearly at $H=7$\,kOe
from a disordered state prevailing below $4.5$\,kOe. Further, at
an elevated field ($15$\,kOe), this ratio decreases substantially
as if a transition from the BG phase into a VG like phase has occurred
there. 

In the end, we draw attention towards the thermomagnetic history effects
investigated at lower fields (Figs.~5-7). It is customary to witness
a more disordered FC state than the ZFC in the vicinity of the PE
phenomenon \cite{key-19,key-20,key-21,key-22}. However, the history-dependent
magnetization behavior at lower fields ($H<H^{*}$) studied here in
Yb$_{3}$Rh$_{4}$Sn$_{13}$ have revealed that the vortex matter
in the ZFC mode is more disordered than that in the FC case. The following
scenario seems plausible to explain this feature. In the ZFC mode,
the injection of vortices at high velocities into a superconducting
specimen is influenced by surface barriers and edge effects \cite{key-46,key-47}
such that the vortices eventually penetrate the sample through the
weakest point of the barrier. Inside the superconducting specimen,
the vortices moving in bundles get randomly pinned at respective pinning
sites such that the inter-vortex spacing is non-unique which leads
to a non-uniform distribution of the flux-density. As mentioned earlier,
the vortices are well separated at lower fields and therefore, the
(elastic) interactions between them remain weak resulting in a stronger
pinned (disordered) vortex configuration in the ZFC mode. Although
the vortex matter created in the FC mode also remains disordered at
lower fields, however, the vortices in this mode nucleate more uniformly
as they remain oblivious to disordering effects of the surface barriers
and the non-uniform injection through edges, etc. Therefore, at lower
fields, the vortex configuration in FC mode is less disordered than
in the case of ZFC. At higher fields, as the vortices get closer to
each other and hence, the interaction effects become dominant yielding
a better spatial ordering even during the moving state of creation
of vortex matter in the ZFC mode. On the other hand, during the FC
mode, the vortex matter has to cross the PE boundary in the $H$--$T$
phase space which results in supercooling the disordered vortex matter
prevailing at the peak position of the PE anomaly. Thus, at higher
fields, the vortex matter created in the ZFC mode is more spatially
ordered than that in the FC mode.

\section{Conclusion}

We have investigated via magnetization measurements, a weakly\,-\,pinned
single crystal of a low $T_{c}$ superconductor, Yb$_{3}$Rh$_{4}$Sn$_{13}$.
The present results have lead to the identification of BG to VG transition
line and sketch of a characteristic field ($H^{*}(T)$) line in the
$H$--$T$ phase space of this compound. Surprisingly, the SMP transition
like phase boundary has been unearthed under the combined influence
of an ac driving force and a continuous magnetic field sweeping involved
in the (isothermal) ac susceptibility ($\chi\thinspace^{\prime}(H)$)
measurements. This transition was, however, not observed in the (isothermal)
dc $M$--$H$ loops and the temperature\,-\,dependent ac susceptibility
scans ($\chi\thinspace^{\prime}(T)$). The latter two modes of measurements
yield signature of only the quintessential PE anomaly signaling the
collapse of elasticity of vortex solid at higher fields. An apparent
demarcation between the domain of SMP like transition and the onset
of PE anomaly has been made in the vortex phase diagram of Yb$_{3}$Rh$_{4}$Sn$_{13}$.
The results presented in our specimen answer in affirmative the question
of generic nature of BG to VG transition in a pinned superconductor.
These also find supports from another experimental study \cite{key-45}
at microscopic level in a sample of the same compound. In the low
field region ($H<H^{*}$), the vortex matter is construed to be highly
disordered. Here, a vortex state created in the FC mode is found to
be more ordered than that obtained in the ZFC mode.
\begin{acknowledgments}
Santosh Kumar would like to acknowledge the Council of Scientific
and Industrial Research, India for the grant of the Senior Research
Fellowship. Santosh Kumar wishes to thank Ulhas Vaidya for his help
and assistance in the use of SVSM system in TIFR, Mumbai in the initial
phase of work.\end{acknowledgments}


\begin{thebibliography}{10}
\bibitem{key-1}T. Giamarchi and P. Le. Doussal, Phys. Rev. Lett.
72, 1530 (1994). 

\bibitem{key-2}T. Giamarchi and P. Le. Doussal, Phys. Rev. B 52,
1242 (1995). 

\bibitem{key-3}S. S. Banerjee, A. K. Grover, M. J. Higgins, Gutam
I. Menon, P. K. Mishra, D. Pal, S. Ramakrishnan, T. V. Chandrasekhar
Rao, G. Ravikumar, V. C. Sahni, S. Sarkar, C. V. Tomy, Physica C 355,
39 (2001) and references therein. 

\bibitem{key-4}T. Giamarchi and P. Le. Doussal, Phys. Rev. B 55,
6577 (1997). 

\bibitem{key-5}M. Daeumling, J. M. Seuntjens and D. C. Larbalestier,
Nature (London) 346, 332 (1990). 

\bibitem{key-6}B. Khaykovich, E. Zeldov, D. Majer, T. W. Li, P. H.
Kes and M. Konczykowski, Phys. Rev. Lett. 76, 2555 (1996). 

\bibitem{key-7}B. Khaykovich, M. Konczykowski, E. Zeldov, R. A. Doyle,
D. Majer, P. H. Kes and T. W. Li, Phys. Rev. B 56, R517 (1997). 

\bibitem{key-8}T. Nishizaki, N. Kobayashi, Supercond. Sci. Technol.
13, 1 (2001). 

\bibitem{key-9}D. Pal, S. Ramakrishnan, A. K. Grover, D. Dasgupta,
B. K. Sarma, Phys. Rev. B 63, 132505 (2001). 

\bibitem{key-10}S. Sarkar, D. Pal, P. L. Paulose, S. Ramakrishnan,
A. K. Grover, C. V. Tomy, D. Dasgupta, Bimal, K. Sarma, G. Balakrishnan,
D. McK Paul, Phys. Rev. B 64, 144510 (2001). 

\bibitem{key-11}D. Giller, B. Kalisky, I. Shapiro, B. Ya. Shapiro,
A. Shaulov, Y. Yeshurun, Physica C 388, 731 (2003). 

\bibitem{key-12}A. B. Pippard, Philos. Mag. 19, 217 (1969). 

\bibitem{key-13}M. J. Higgins, S. Bhattacharya, Physica C 257, 232
(1996) and references therein. 

\bibitem{key-14}H. Sato, Y. Aoki, H. Sugawara, T. Fukuhara, J. Phys.
Soc. Jpn. 64, 3175 (1995). 

\bibitem{key-15}C. V. Tomy, G. Balakrishnan, D. McK. Paul, Physica
C 280, 1 (1997). 

\bibitem{key-16}S. Sarkar, S. Banerjee, A. K. Grover, S. Ramakrishnan,
S. Bhattacharya, G. Ravikumar, P. K. Mishra, V. C. Sahni, C. V. Tomy,
D. McK. Paul, G. Balakrishnan, M. J. Higgins, Physica C 341, 1055
(2000). 

\bibitem{key-17}X. S. Ling, S. R. Park, B. A. McClain, S. M. Choi,
D. C. Dender, J. W. Lynn, Phys. Rev. Lett. 86, 712 (2001). 

\bibitem{key-18}Santosh Kumar, Ravi P. Singh, A. Thamizhavel, C.
V. Tomy and A. K. Grover, Physica C 506, 69 (2014). 

\bibitem{key-19}W. Henderson, E. Y. Andrei, M. J. Higgins and S.
Bhattacharya, Phys. Rev. Lett. 77, 2077 (1996). 

\bibitem{key-20}S. S. Banerjee, N. G. Patil, S. Ramakrishnan, A.
K. Grover, S. Bhattacharya, G. Ravikumar, P. K. Mishra, T. V. Chandrasekhar
Rao, V. C. Sahni and M. J. Higgins, Appl. Phys. Lett. 74, 126 (1999). 

\bibitem{key-21}S. S. Banerjee, N. G. Patil, S. Ramakrishnan, A.
K. Grover, S. Bhattacharya, P. K. Mishra, G. Ravikumar, T. V. Chandrasekhar
Rao, V. C. Sahni, M. J. Higgins, C. V. Tomy, G. Balakrishnan and D.
McK Paul, Phys Rev. B 59, 6043 (1999). 

\bibitem{key-22}G. Ravikumar, V. C. Sahni, P. K. Mishra, T. V. Chandrasekhar
Rao, S. S. Banerjee, A. K. Grover, S. Ramakrishnan, S. Bhattacharya,
M. J. Higgins, E. Yamamoto, Y. Haga, M. Hedo and Y. Inada, Y. Onuki,
Phys. Rev. B 57, R11069 (1998). 

\bibitem[23]{key-23}G. P. Espinosa, Mater. Res. Bull. 15, 791 (1980). 

\bibitem[24]{key-24}C.P. Bean, Rev. Mod. Phys. 36, 31 (1964). 

\bibitem[25]{key-25}W.A. Fietz and W.W. Webb, Phys. Rev. 178, 657
(1969).

\bibitem[26]{key-26}M. Marchevsky, M. J. Higgins and S. Bhattacharya,
Nature 409, 591 (2001). 

\bibitem[27]{key-27}X. S. Ling and J. Budnick in Magnetic Susceptibility
of Superconductors and Other Spin Systems, edited by R. A. Hein, T.
L. Francavilla, D. H. Liebenberg (Plenum Press, New York, 1991), p.
377. 

\bibitem[28]{key-28}Y. Paltiel, E. Zeldov, Y. Myasoedov, M. L. Rappaport,
G. Jung, S. Bhattacharya, M. J. Higgins, Z. L. Xiao, E. Y. Andrei,
P. L. Gammel and D. J. Bishop, Phys. Rev. Lett. 85, 3712 (2000). 

\bibitem[29]{key-29}G. Ravikumar, H. Kupfer, A. Will, R. Meier-Hirmer
and Th. Wolf, Phys. Rev. B 65, 094507, (2002). 

\bibitem[30]{key-30}A.I. Larkin and Yu. N. Ovchinnikov, Sov. Phys.
JETP 38, 854 (1974). 

\bibitem[31]{key-31}A.I. Larkin and Yu. N. Ovchinnikov, J. Low Temp.
Phys. 34, 409 (1979). 

\bibitem[32]{key-32}John R. Clem and Zhidong Hao, Phys. Rev. B 48,
13774 (1993). 

\bibitem[33]{key-33}S. O. Valenzuela and V. Bekeris, Phys. Rev. Lett.
84, 4200 (2000). 

\bibitem[34]{key-34}G. Ravikumar, K. V. Bhagwat, V. C. Sahni, A.
K. Grover, S. Ramakrishnan and S. Bhattacharya, Phys. Rev. B 61, R6479
(2000). 

\bibitem[35]{key-35}G. Ravikumar, V. C. Sahni, A. K. Grover, S. Ramakrishnan,
P. L. Gammel, D. J. Bishop, E. Bucher, M. J. Higgins and S. Bhattacharya,
Phys. Rev. B 63, 024505 (2001). 

\bibitem[36]{key-36}G. Ravikumar and H Kupfer, Phys. Rev. B 72, 144530
(2005). 

\bibitem[37]{key-37}N. Avraham, B. Khaykovich, Y. Myasoedov, M. Rappaport,
H. Shtrikman, D. E. Feldman, T. Tamegai, P. H. Kes, M. Li, M. Konczykowski,
K. van der Beek, E. Zeldov, Nature 411, 451 (2001). 

\bibitem[38]{key-38}S. S. Banerjee \textit{et al.}, Physica C 308,
25-32 (1998). 

\bibitem[39]{key-39}D. S. Fisher, M. P. A. Fisher, and D. A. Huse,
Phys. Rev. B 43, 130 (1991).. 

\bibitem[40]{key-40}K. Ghosh, S. Ramakrishnan, A. K. Grover, Gautam
I. Menon, Girish Chandra, T. V. Chandrasekhar Rao, G. Ravikumar, P.
K. Mishra, V. C. Sahni, C. V. Tomy, G. Balakrishnan, D. McK Paul,
S. Bhattacharya, Phys. Rev. Lett. 76, 4600 (1996). 

\bibitem[41]{key-41}M. J. P. Gingras and D. A. Huse, Phys. Rev. B
53, 15193 (1996).

\bibitem[42]{key-42}D. Pal, D. Dasgupta, Bimal K Sarma, S. Bhattacharya,
S. Ramakrishnan and A. K. Grover, Phys. Rev. B 62, 6699 (2000). 

\bibitem[43]{key-43}D.G. Grier, C.A. Murray, C.A. Bolle, P.L. Gammel,
D.J. Bishop, D.B. Mitzi, and A. Kapitulnik, Phys. Rev. Lett. 66, 2270
(1992).

\bibitem[44]{key-44}S. Horiuchi, M. Cantoni, M. Uchida, T. Tsuruta,
and Y. Matsui, Appl. Phys. Lett. 73, 1293 (1998).

\bibitem[45]{key-45}D. Mazzone, J. L. Gavilano, R. Sibille, M. Ramakrishnan,
C. D. Dewhurst, and M. Kenzelmann, arXiv:1407.0569v2 {[}cond-mat.supr-con{]}
22 Oct 2014. 

\bibitem[46]{key-46}Y. Paltiel, D. T. Fuchs, E. Zeldov, Y. N. Myasoedov,
and H. Shtrikman, M. L. Rappaport and E. Y. Andrei, Phys. Rev. B 58,
R14763 (1998).

\bibitem[47]{key-47}Y. Paltiel, E. Zeldov, Y. N. Myasoedov, H. Shtrikman,
S. Bhattacharya, M. J. Higgins, Z. L. Xiao, E. Y. Andrei, P. L. Gammel
and D. J. Bishop, Nature (London) 403, 398 (2000). \end{thebibliography}
\end{document}